# Some results on communicating the sum of sources over a network

Brijesh Kumar Rai, Bikash Kumar Dey, and Abhay Karandikar
Department of Electrical Engineering
Indian Institute of Technology Bombay
Mumbai, India, 400 076
{bkrai,bikash,karandi}@ee.iitb.ac.in

*Abstract*—We consider the problem of communicating the sum of $m$ sources to $n$ terminals in a directed acyclic network. Recently, it was shown that for a network of unit capacity links with either $m = 2$ or $n = 2$, the sum of the sources can be communicated to the terminals if and only if every source-terminal pair is connected in the network. We show in this paper that for any finite set of primes, there exists a network where the sum of the sources can be communicated to the terminals only over finite fields of characteristic belonging to that set. As a corollary, this gives networks where the sum can not be communicated over any finite field even though every source is connected to every terminal.

## I. INTRODUCTION

The seminal work by Ahlswede et al. [1] showed that in a multicast network, higher rates are achievable if the intermediate nodes in the network perform some processing of the incoming information before forwarding. This has been popularly known as network coding since then. Even though the multicast capacity of a network under routing is difficult to compute or characterize in general, the authors showed in [1] that under network coding, the multicast capacity is given by the minimum of the min-cut capacities of the individual terminals. Li et al. [2] showed that the multicast capacity is also achievable by linear network coding, i.e., with each intermediate node computing linear combinations of the incoming messages for transmitting on the outgoing links. An algebraic formulation was presented in [3]. It was also shown that linear network codes can be designed to be robust against link failure. Jaggi et al. [4] gave a polynomial time algorithm for designing multicast network codes.

Recently, the problem of communicating the sum of sources to some terminals was considered by Ramamoorthy [5]. It was shown that if there are two sources or two terminals in the network, then the sum of the sources can be communicated to the terminals if and only if every source is connected to every terminal. Whereas this condition is also necessary for any number of sources and terminals, it may not be sufficient. Further, no necessary and sufficient condition is known for arbitrary number of sources and terminals.

The problem of distributed function computation in general has been considered in different contexts in the past. Distributed event detection and data aggregation techniques in sensor network has been of significant interest [6], [7] since the availability of cheap efficient sensors. Distributed computation of the sum/parity of the binary sources in a large network was first considered by Gallager [8]. There has been significant interest in computation of such functions over random geometric graphs motivated by wireless sensor network, e.g., [9], [10]. There is also significant interest in the computation of such functions over arbitrary network graphs, e.g., [11].

In this paper, we consider a directed acyclic network. We show that for every finite set of prime numbers, there exists a directed acyclic network of unit-capacity links with some sources and terminals so that the sum of the sources can be communicated to all the terminals if and only if the characteristic of the alphabet field belongs to the given set. As a corollary, we find a network of 3 sources and 3 terminals where the sum of the sources can not be communicated to the terminals over any finite field even though every source is connected to every terminal in the network. This example network was also independently found by Ramamoorthy and Langberg [12].

In Section II, we introduce the system model. The results of this paper are presented in Section III. We conclude the paper with a discussion in Section IV.

## II. SYSTEM MODEL

The network is represented by a directed acyclic graph $G = (V, E)$ where $V$ is a finite set denoting the vertices of the network, $E \subseteq V \times V$ is the set of edges. Among the vertices, there are $m$ sources $s_1, s_2, \cdots, s_m \in V$, and $n$ terminals $t_1, t_2, \cdots, t_n \in V$ in the network. In general, each terminal node may have a requirement of recovering some part of the source messages or their functions. We consider a network where each terminal node requires to recover the sum of the source messages. For any edge $e = (i, j) \in E$, the node $j$ will be called the head of the edge and the node $i$ will be called the tail of the edge; and they will be denoted as $head(e)$ and $tail(e)$ respectively. Throughout the paper, $p$, possibly with subscripts, will denote a positive prime integer, and $q$ will denote a power of a prime. Let $\mathbb{F}_q$ denote the alphabet field. Each link in the network is assumed to be capable of carrying a symbol from $\mathbb{F}_q$ in each use. Each symbol interval uses the channel once and this time is taken as the unit time.

For any edge $e \in E$, let $Y_e \in \mathbb{F}_q$ denote the message transmitted through $e$. In scalar linear network coding, each node computes a linear combination of the incoming symbols for transmission on an outgoing link. That is,

$$Y_e = \sum_{e':head(e')=tail(e)} \beta_{e',e} Y_{e'} \qquad (1)$$

when $tail(e)$ is not a source node. Here $\beta_{e',e} \in \mathbb{F}_q$ are called the local coding coefficients. A source node computes a linear combination of some data symbols generated at that source for transmission on an outgoing link, that is,

$$Y_e = \sum_{j: X_j \text{ generated at } tail(e)} \alpha_{j,e} X_j \qquad (2)$$

for some $\alpha_{j,e} \in \mathbb{F}_q$ if $tail(e)$ is a source node. We assume that each source generates one symbol from $\mathbb{F}_q$ per unit time. So, there is only one term in the summation in (2) and $\alpha_{j,e}$ can be taken to be 1 without loss of generality. The decoding operation at a terminal involves taking a linear combination of the incoming messages to recover the required data.

In vector linear network coding, the data stream generated at each source node is blocked in vectors of length $N$. The coding operations are similar to Eq. (1) and Eq. (2) with the difference that, now $Y_e, Y_{e'}, X_j$ are vectors from $\mathbb{F}_q^N$, and $\beta_{e',e}, \alpha_{j,e}$ are matrices from $\mathbb{F}_q^{N \times N}$. It is known that scalar linear network coding may give better throughput in some networks than that is achievable by routing. Vector linear network coding may give further improvement over scalar linear network coding in some networks [13], [14], [15].

## III. RESULTS

A complete bipartite graph $K_{m,n}$ has $m$ nodes in one partition and $n$ nodes in the other. We will assume that the $m$ nodes in one partition are sources and the $n$ nodes in the other partition are the terminals in the corresponding network. All the edges are assumed to be directed from the source partition to the terminal partition. Clearly, in the network $K_{m,n}$, each source node can broadcast its message to all the terminals and each terminal node can thus recover any function of the source messages. In particular, each terminal node can recover the sum of the source messages.

We now define a special class of networks.

We define a network $\mathcal{S}_m \triangleq (V(\mathcal{S}_m), E(\mathcal{S}_m))$ which has four layers of vertices $V(\mathcal{S}_m) = S \cup U \cup V \cup T$. The first layer of nodes are the $m$ source nodes $S \triangleq \{s_1, s_2, \ldots, s_m\}$. The second and third layers have $m-1$ nodes each, and they are denoted as $U \triangleq \{u_1, u_2, \ldots, u_{m-1}\}$ and $V \triangleq \{v_1, v_2, \ldots, v_{m-1}\}$ respectively. The last layer consists of the $m$ terminal nodes $T \triangleq \{t_1, t_2, \ldots, t_m\}$. For every $i = 1, 2, \ldots, m-1$, there is an edge from $s_i$ to $u_i$, from $u_i$ to $v_i$, and from $v_i$ to $t_i$. That is, $(s_i, u_i), (u_i, v_i), (v_i, t_i) \in E(\mathcal{S}_m)$ for each $i = 1, 2, \ldots, m-1$. For every $i, j = 1, 2, \ldots, m-1$, $i \neq j$, there is an edge from $s_i$ to $t_j$. Finally, for every $i = 1, 2, \ldots, m-1$, there is an edge from $s_m$ to $u_i$ and from $v_i$ to $t_m$. So, the set of edges is given by

$$\begin{aligned}
E(\mathcal{S}_m) = &\cup_{i=1}^{m-1} \{(s_i, u_i), (u_i, v_i), (v_i, t_i)\} \\
&\cup \{(s_i, t_j) : i, j = 1, 2, \ldots, m-1, i \neq j\} \\
&\cup \{(s_m, u_i) : i = 1, 2, \ldots, m-1\} \\
&\cup \{(v_i, t_m) : i = 1, 2, \ldots, m-1\}
\end{aligned}$$

The network is shown in Fig. 1.

Now we define a method of combining two networks to obtain a larger network. We call this method *criss-crossing*. Let $\mathcal{N}_1$ be a directed acyclic network with some source nodes $S_1 \subseteq V(\mathcal{N}_1)$ and some terminal nodes $T_1 \subseteq V(\mathcal{N}_1)$. Similarly let $\mathcal{N}_2$ be a directed acyclic network with some source nodes $S_2 \subseteq V(\mathcal{N}_2)$ and some terminal nodes $T_2 \subseteq V(\mathcal{N}_2)$. We assume that the nodes of $\mathcal{N}_1$ and $\mathcal{N}_2$ are labeled such that $V(\mathcal{N}_1) \cap V(\mathcal{N}_2) = \phi$.

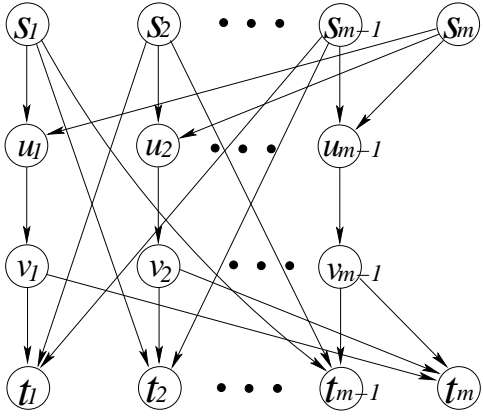

Fig. 1. The network $\mathcal{S}_m$

The crisscrossed network $\mathcal{N}_1 \bowtie \mathcal{N}_2$ has the node set $V(\mathcal{N}_1 \bowtie \mathcal{N}_2) = V(\mathcal{N}_1) \cup V(\mathcal{N}_2)$, and the edge set $E(\mathcal{N}_1 \bowtie \mathcal{N}_2) = E(\mathcal{N}_1) \cup E(\mathcal{N}_2) \cup (S_1 \times T_2) \cup (S_2 \times T_1)$. That is, other than the edges of $\mathcal{N}_1$ and $\mathcal{N}_2$, their crisscross has edges from the sources of $\mathcal{N}_1$ to the terminals of $\mathcal{N}_2$, and from the sources of $\mathcal{N}_2$ to the terminals of $\mathcal{N}_2$. The crisscross of the networks $\mathcal{S}_4$ and $K_{2,3}$ is shown in Fig. 2 for example.

Now we present our main result.

*Theorem 1:* For any finite set $\mathcal{P} = \{p_1, p_2, \ldots, p_l\}$ of positive prime numbers, there exists a directed acyclic network of unit-capacity edges where it is possible to communicate the sum of the source messages at rate one with linear network coding if and only if the characteristic of the alphabet field belongs to $\mathcal{P}$.

**Proof:** Define $m = p_1 p_2 \ldots p_l + 2$. We prove that the network $\mathcal{S}_m$ satisfies the condition in the theorem. First, it may be noted that every source-terminal pair in the network $\mathcal{S}_m$ is connected. This is clearly a necessary condition for being able to communicate the sum of the source messages to each terminal node over any field.

First we prove that if it is possible to communicate the sum of the source messages by vector linear network coding over $\mathbb{F}_q$ to all the terminals in $\mathcal{S}_m$, then the characteristic of $\mathbb{F}_q$ must be from $\mathcal{P}$. As in Eq. (1) and Eq. (2), the message carried by an edge $e$ is denoted by $Y_e$. For $i = 1, 2, \ldots, m$, the message vector generated by the source $s_i$ is denoted by $X_i \in \mathbb{F}_q^N$. Each terminal $t_i$ computes a linear combination $R_i$ of the received vectors.

Local coding coefficients used at different layers in the network are denoted by different symbols for clarity. The message vectors carried by different edges and the corresponding local coding coefficients are as below.

$$Y_{(s_i,t_j)} = \alpha_{i,j} X_i \text{ for } 1 \leq i,j \leq m-1, i \neq j \quad (3a)$$
$$Y_{(s_i,u_i)} = \alpha_{i,i} X_i \text{ for } 1 \leq i \leq m-1 \quad (3b)$$
$$Y_{(s_m,u_i)} = \alpha_{m,i} X_m \text{ for } 1 \leq i \leq m-1 \quad (3c)$$
$$Y_{(u_i,v_i)} = \beta_{i,1} Y_{(s_i,u_i)} + \beta_{i,2} Y_{(s_m,u_i)}$$
$$\text{for } 1 \leq i \leq m-1 \quad (3d)$$

$$R_i = \sum_{\substack{j=1 \\ j \neq i}}^{m-1} \gamma_{j,i} Y_{(s_j,t_i)} + \gamma_{i,i} Y_{(v_i,t_i)}$$
$$\text{for } 1 \leq i \leq m-1 \quad (4a)$$
$$R_m = \sum_{j=1}^{m-1} \gamma_{j,m} Y_{(v_j,t_m)}. \quad (4b)$$

Here all the coding coefficients $\alpha_{i,j}, \beta_{i,j}, \gamma_{i,j}$ are $N \times N$ matrices over $\mathbb{F}_q$, and the message vectors $X_i$ and the messages carried by the links $Y_{(\cdot,\cdot)}$ are length-$N$ vectors over $\mathbb{F}_q$.

Without loss of generality (w.l.o.g.), we assume that
$$Y_{(v_i,t_i)} = Y_{(v_i,t_m)} = Y_{(u_i,v_i)}.$$

By assumption, each terminal decodes the sum of all the source messages. That is,
$$R_i = \sum_{j=1}^{m} X_j \text{ for } i = 1, 2, \ldots, m \quad (5)$$

for all values of $X_1, X_2, \ldots, X_m \in \mathbb{F}_q^N$.

From equations (3) and (4), we have
$$R_i = \sum_{\substack{j=1 \\ j \neq i}}^{m-1} \gamma_{j,i} \alpha_{j,i} X_j + \gamma_{i,i} \beta_{i,1} \alpha_{i,i} X_i + \gamma_{i,i} \beta_{i,2} \alpha_{m,i} X_m \quad (6)$$

for $i = 1, 2, \ldots, m-1$, and
$$R_m = \sum_{j=1}^{m-1} \gamma_{j,m} \beta_{j,1} \alpha_{j,j} X_j + \sum_{j=1}^{m-1} \gamma_{j,m} \beta_{j,2} \alpha_{m,j} X_m. \quad (7)$$

Since (5) is true for all values of $X_1, X_2, \ldots, X_m \in \mathbb{F}_q^N$, equations (6) and (7) imply
$$\gamma_{j,i} \alpha_{j,i} = I \text{ for } 1 \leq i,j \leq m-1, i \neq j \quad (8)$$
$$\gamma_{i,i} \beta_{i,1} \alpha_{i,i} = I \text{ for } 1 \leq i \leq m-1 \quad (9)$$
$$\gamma_{i,i} \beta_{i,2} \alpha_{m,i} = I \text{ for } 1 \leq i \leq m-1 \quad (10)$$
$$\gamma_{i,m} \beta_{i,1} \alpha_{i,i} = I \text{ for } 1 \leq i \leq m-1 \quad (11)$$
$$\sum_{i=1}^{m-1} \gamma_{i,m} \beta_{i,2} \alpha_{m,i} = I \quad (12)$$

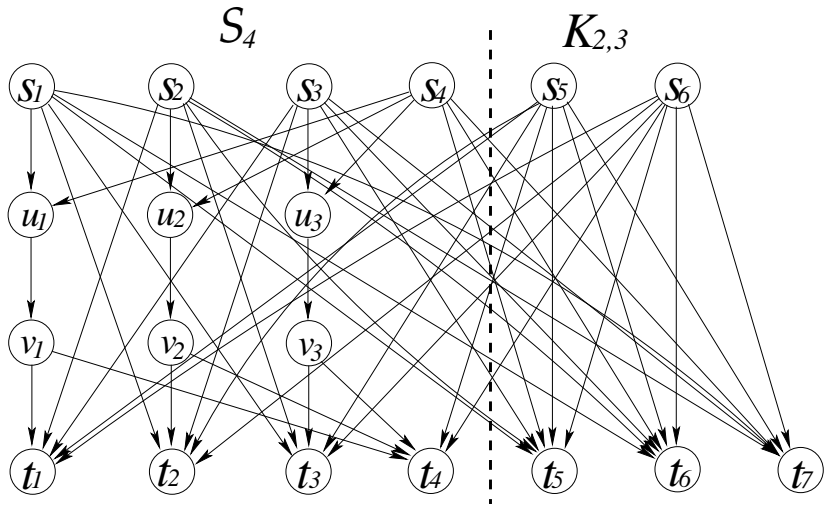

Fig. 2. The network $\mathcal{S}_4 \bowtie K_{2,3}$

where $I$ denotes the $N \times N$ identity matrix over $\mathbb{F}_q$. All the coding matrices in equations (8),(9),(10), (11) are invertible since the right hand side of the equations are the identity matrix. Equations (9) and (10) imply $\beta_{i,1}\alpha_{i,i} = \beta_{i,2}\alpha_{m,i}$ for $1 \leq i \leq m-1$. So, (12) gives

$$\sum_{i=1}^{m-1} \gamma_{i,m}\beta_{i,1}\alpha_{i,i} = I \qquad (13)$$

Now, using (11), we get

$$\sum_{i=1}^{m-1} I = I \qquad (14)$$
$$\Rightarrow (m-1)I = I \qquad (15)$$
$$\Rightarrow (m-2)I = \mathbf{0}. \qquad (16)$$

This is true if and only if the characteristic of $\mathbb{F}_q$ divides $m-2$. So, the sum of the source can be communicated in $\mathcal{S}_m$ by vector linear network coding at rate 1 only if the characteristic of $\mathbb{F}_q$ belongs to $\mathcal{P}$.

Now, if the characteristic of $\mathbb{F}_q$ belongs to $\mathcal{P}$, then for any block length $N$, in particular for scalar network coding for $N = 1$, every coding matrix in (3a)-(3d) can be chosen to be the identity matrix. The terminals then can recover the sum of the source messages by taking the sum of the incoming messages, i.e., by taking $\gamma_{j,i} = I$ for $1 \leq j \leq m-1$ and $1 \leq i \leq m$ in (4a) and (4b).

When the set $\mathcal{P}$ is a singleton, Theorem 1 gives, as a special case, the following corollary.

*Corollary 2:* For any prime number $p$, the sum of the source messages can be communicated to all the terminals by vector linear network coding in the network $\mathcal{S}_{p+2}$ only over fields of characteristic $p$. Moreover, over fields of characteristic $p$, the sum can be communicated in this network by scalar network coding.

When the set $\mathcal{P}$ is empty, Theorem 1 gives, as a special case, a network where the the sum of the source messages can not be communicated using linear network coding over any field.

*Corollary 3:* In the network $\mathcal{S}_3$, the sum of the sources can not be communicated to the terminals over any finite field.

It was proved in [5] that if $m < 3$ or $n < 3$, then any network where every source-terminal pair is connected allows the sum of the sources to be communicated to the terminals by linear network coding. The network $\mathcal{S}_3$, shown in Fig. 3, is an example of a network with $m, n \geq 3$ where the sum of the sources can not be communicated to the terminals by linear network coding even though every source-terminal pair is connected. This example was also found independently by Rammoorthy and Langberg [12].

Let $\mathcal{P}(\mathcal{N})$ denote the set of characteristics of fields over which the sum of the sources can be communicated to the terminals by linear network coding in the network $\mathcal{N}$. We have the following results.

*Theorem 4:* For any two networks $\mathcal{N}_1$ and $\mathcal{N}_2$,

$$\mathcal{P}(\mathcal{N}_1 \bowtie \mathcal{N}_2) = \mathcal{P}(\mathcal{N}_1) \cap \mathcal{P}(\mathcal{N}_2).$$

**Proof:** For any two networks $\mathcal{N}_1$ and $\mathcal{N}_2$, and for any field $F$, it is possible to communicate the sum of the sources to the terminals in the crisscrossed network $\mathcal{N}_1 \bowtie \mathcal{N}_2$ by linear network coding over $F$ if and only if it is possible to communicate the sum of the sources to the terminals in the individual networks $\mathcal{N}_1$ and $\mathcal{N}_2$

by linear network coding over $F$. So the result follows. □

*Corollary 5:* For any network $\mathcal{N}$, and any positive integers $m, n$,

$$\mathcal{P}(\mathcal{N} \bowtie K_{m,n}) = \mathcal{P}(\mathcal{N}).$$

For any positive integer $m$, let $\Pi(m)$ denote the set of prime factors of $m$. Then Theorem 1 states that $\mathcal{P}(\mathcal{S}_m) = \Pi(m-2)$. Theorem 4 then gives

*Corollary 6:* For any two positive integers $m$ and $n$,

$$\mathcal{P}(\mathcal{S}_m \bowtie \mathcal{S}_n) = \Pi(\gcd(m-2, n-2))$$

Theorem 4 together with the two corollaries allow construction of a large class of networks having arbitrary finite $\mathcal{P}(\mathcal{N})$ and arbitrary size. For example, for any positive $m, n > 3$, one can construct a network by crisscrossing $\mathcal{S}_3$ with an appropriate complete bipartite network to get a network with $m$ sources and $n$ terminals where the sum of the sources can not be communicated to the terminals by linear network coding over any field. Such networks can also be constructed by crisscrossing networks with disjoint $\Pi$. For instance, for any two prime numbers $p_1$ and $p_2$, the sum of the sources can not be communicated to the terminals in the network $\mathcal{S}_{p_1+2} \bowtie \mathcal{S}_{p_2+2}$.

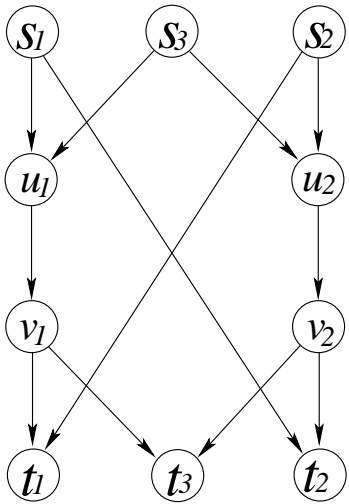

Fig. 3. The network $\mathcal{S}_3$

## IV. Discussion

We have constructed networks where the sum of the sources can be communicated by linear network coding only over fields of a specified finite set of characteristics. A necessary and sufficient condition for being able to communicate the sum of the sources over a field to all the terminals is still not known. Such a condition for fields of both zero and non-zero characteristic are of interest. Communication of the sum over fields like $\mathbb{R}$ and $\mathbb{C}$ of zero characteristic fields are of interest, for example, in sensor network when the real measurement values are transmitted as analog values. Many other functions of the source messages may also be of practical interest.

## V. Acknowledgments

This work was supported in part by Tata Teleservices IIT Bombay Center of Excellence in Telecomm (TICET).